\documentclass[twocolumn,twoside,slac_two]{revtex4}
\usepackage{graphicx}
\usepackage{fancyhdr}
\begin{document}

\title{Consequences of the LHC Results in the Interpretation of $\gamma$ ray 
families and Giant EAS Data}

%

\author{J. N. Capdevielle}
\affiliation{APC, Univ.Paris Diderot, 10 rue A.Domon, 75205 Paris, France}
%

\begin{abstract}
The earliest results of CMS exhibit central pseudo rapidity densities larger than the predictions of the different models. Introducing on this basis new guidelines with larger multiplicities of secondaries in the models implemented in the simulations, we examine the consequences in $\gamma$ ray families (spikes in rapidity distribution, coplanar emission) and very large EAS (penetration power in the atmosphere).
\\

\end{abstract}

\maketitle

\thispagestyle{fancy}


\section{INTRODUCTION}
The collection of high energy cosmic ray events performed with X ray emulsion chambers (XREC) in the energy range
covered by the LHC has been the origin of questions for high energy physics.  Conversely, the experience of the colliders and  the earliest results of the LHC help considerably the interpretation of those events by sophisticated simulations;  the separation of genuine hints of new physics among complex effects of natural fluctuations becomes more reliable. Usually, the gamma ray families recorded with XREC at very high altitudes are not contained events;
 the origin of the collision can  be only approached by the invariant mass of the $\pi_{0}$'s and sometimes by the geometry of the most energetic subcascades.
   
Aside from those events which are more or less dependent on the primary composition, the extensive air shower data, through the characteristic properties of the different components is also correlated with the important features of
 the primary collisions up to energies larger by 3 orders of magnitude (in laboratory system) 
 than the present limit of the LHC.
Additive  uncertainties come from  the cosmic ray particles interaction with the air nuclei (N, O, Ar) instead of nucleons, the lower energy limits of the experiments in p-A collisions  and a lack of observation in the fragmentation region in colliders. 
\section{REMARKABLE COSMIC RAY EVENTS}
The Centauros have been widely studied in the previous ISVHECRI symposia and measurements are done in colliders in relation with $N_{\gamma}-N_{ch}$ relation.
Hence, this section is  concentrated on other striking features such as the abnormal fluctuations in pseudo rapidity distribution and the coplanar emission.
The observation of the pseudo rapidity distribution near an energy above $10^{6}$ GeV is rare ; we may quote the Texas Lone Star, a dozen of A-A collisions recorded in balloons by the JACEE and an event recorded in the Concorde experiment \cite{cap} exhibiting spikes in the pseudo rapidity distribution (fig.1).
 The localisation of the vertex in the cabin wall of the Concorde airliner  by both triangulation and invariant mass method allowed the reconstruction of this distribution, also plotted in the Centre of Mass System (CMS) on fig.2 assuming a primary energy of $10^{6}$ GeV. The forward cone (142 $\gamma$'s over 149) is compared here with the charged inclusive and semi inclusive NSD pseudo rapidity distributions for $Z_{kno} =2$ and $Z_{kno} =3$ ($Z_{kno} = N_{ch}/<N_{ch}>$ is the Koba-Nielsen-Olesen parameter) averaged over 3000 events generated with our Monte Carlo generator (see section 3).     
 \begin{figure}                                                                  \includegraphics[width=0.5\textwidth,clip]{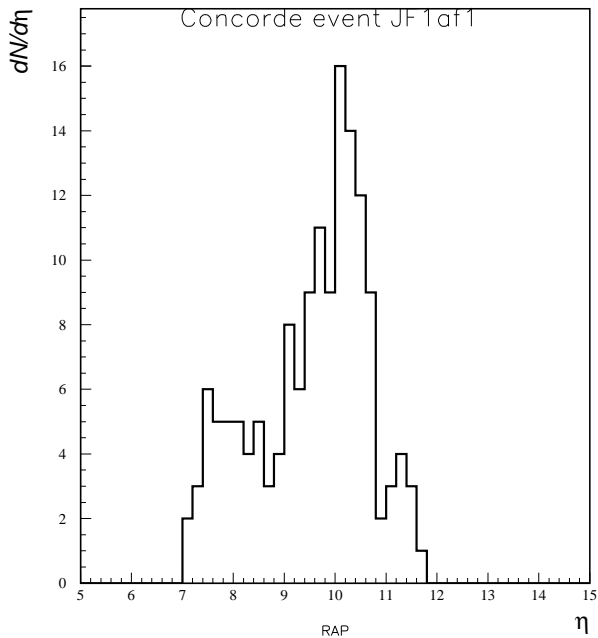} \hfill                    \caption{$\gamma$'s pseudo rapidity distribution in Laboratory System for 149 $\gamma$'s recorded in the event JF1af1 at CONCORDE flight altitude of $17$km ; typical spikes may be observed near 7.6, 10.3 and 11.4 units of rapidity. }                   \end{figure}    
    
The question occurred how those spikes could arise of simple random fluctuations or might indicate the formation of hot spots of quark-gluon plasma \cite{faes}, the peaks of fig.1 being considered as the drops of hadronic matter into which the quark-gluon-plasma condenses.

 \begin{figure}
 \includegraphics[width=0.5\textwidth,clip]{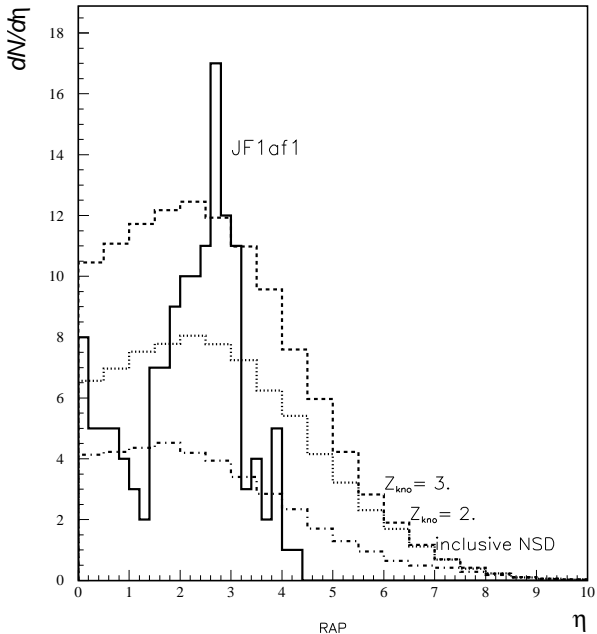} \hfill
 \caption{$\gamma$'s pseudo rapidity distribution in CMS for  142 $\gamma$'s (solid line) recorded in the event JF1af1 ; the spikes are shifted near 2.8 and 3.9 units of rapidity. Simulated NSD distributions are superimposed for the inclusive component (dotted-dashed)and for semi inclusive data with $Z_{kno} = 2$ (dotted) and $Z_{kno} = 3$ (dashed) } 
\end{figure}

Such events have been also searched in UA5 CERN experiment \cite{aln} : the  random grouping of clusters in Monte Carlo simulated events suggest that there could be no need to include here new phenomena. In the case of JF1af1 which is a p-Al collision, a similar investigation might progress by the analysis of individual  $\gamma$'s distributions between  $Z_{kno} = 2$  and $Z_{kno} = 3$ for events corresponding to fig.2.

At energies around $10$ PeV, alignments of $\gamma$'s and hadrons have been observed, suggesting a coplanar emission of the most energetic secondaries. One pair of events in addition to a dozen clear events recorded at mountain altitude have been collected under a minimal cascading in the low stratosphere, JF2af2, the most energetic event during the exposures on the Concorde and Strana during Siberian balloon flights.
JF2af2 is characterised by a near perfect alignment of all the cascades with energy exceeding 50 TeV (fig.3). After extensive simulations, we have shown that alignments may appear in emulsion chamber and result from rare fluctuations of standard collisions \cite{cap1}.
 \begin{figure}                                                                  \includegraphics[width=0.5\textwidth,clip]{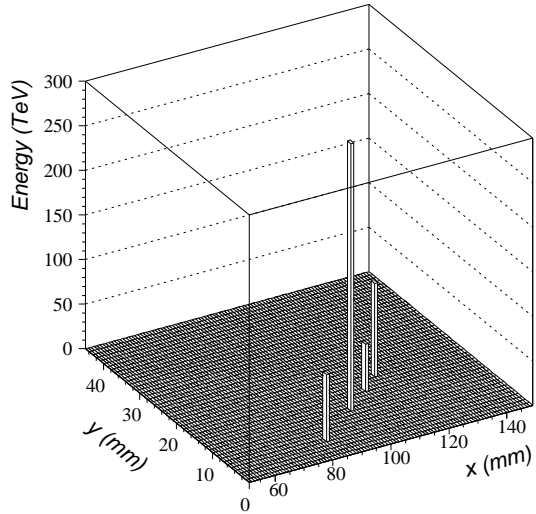} \hfill                    \caption{ Most energetic secondaries among the 211 $\gamma$'s recorded in the event JF2af2. $50\%$ of the visible energy $\sum E_{\gamma}$ ($1586$ TeV) is contained in the alignment.}
                                       \end{figure}     
However, both events JF2af2 and STRANA with a primary energy near $\sqrt(s) = 5 $~TeV exhibit a similar visible energy, large multiplicities 
 and require very large transverse momenta (around $10$~GeV/c).

 The very modest  exposure of the chambers ($500$ hours with an area of $0.2$~m$^{2}$ combined with the important reduction of the primary flux after the knee) objects the hypothesis on standard fluctuations. Therefore, a mechanism involving the valence quarks and diquarks, based on relativistic string fragmentation has been proposed \cite{cap1}:  one  pair $q-\bar{q}$ or $qq-\bar{qq}$ is created when the distance $L$ separating both valence quarks exceeds a threshold value. The string fragmentation tension corresponds to a tension $\kappa = (1/2) \pi \alpha'$ of about 1~GeV/fm, $\alpha'$ being the Regge slope. The transverse momentum of the quarks emitted is related to the tension by the relation:

\begin{equation}
 \sqrt{\langle p_{\rm t} \rangle^2} = \sqrt{\frac{\kappa}{\pi}}
\end{equation}

 The fragmentation of the  new strings stretched between valence quarks and diquarks and also between the partners of the valence diquarks (assumed to be strongly bounded) is the origin of the large $p_{t}$'s : a strong tension of those last  strings ($\kappa$ multiplied by $10-15$) provides large $p_{t}$'s after a violent diquark breaking as in fig.4 for example where $<p_{t}> = 11$ GeV/c$>$, able to reproduce JF2af2. We have here applied the Schwinger mechanism \cite{won}.
 \begin{figure}
 \includegraphics[width=0.5\textwidth,clip]{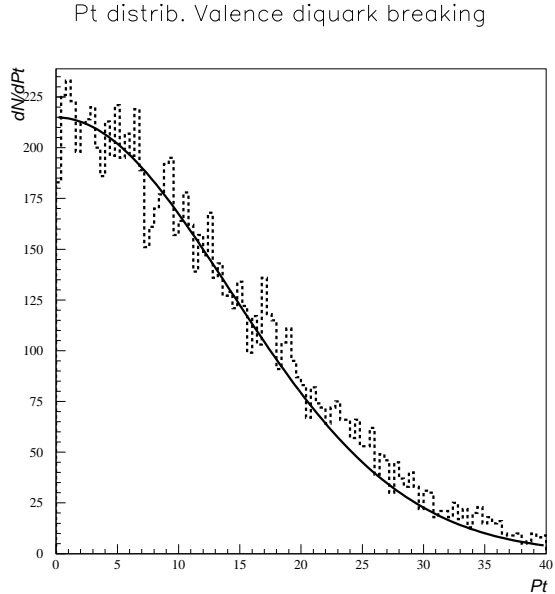} \hfill
 \caption{ $p_{t}$ distribution of valence quarks  assuming valence diquark breaking and Schwinger mechanism.}
\end{figure}

 The fragmentation of such strings would occur above an energy threshold of 200 GeV in CMS per valence quark. Near this threshold the most energetic secondaries are aligned (minimal energy required for maximal tension of the string when 3 valence quarks are aligned); this could explain why the coplanar emission is more likely observed close to $10^{7}$ GeV.
Such configuration is no more necessary at energies exceeding this threshold, the excitation energy being widely sufficient to generate the valence diquark breaking.

  There are some super families at energies above the energy of the LHC (Andromeda, Tadjikistan, Fianit...). Those events present halos, multiple clustering, large multiplicities and large $p_{t}$, but up to now no coplanar emission.      
\section{SIMULATION AND EAS DATA}

\subsection{New Model Guidelines}
Taking the opportunity of the earliest results of the LHC given by CMS \cite{cms}, we begin to update and improve the collision generator HDPM implemented in CORSIKA  and report here the preliminary results obtained with the non-diffractive code.
The parameters of the main guidelines such as the average central pseudo rapidity density $<dN_{ch}/d\eta>$, the average  multiplicity $<N_{ch}>$ have been tuned on the experimental data as follows for $\sqrt(s) > 0.9$ TeV:
\begin{equation}
  <dN_{ch}/d\eta> = 0.595 s^{0.13}
 ~~ <N_{ch}>       = 2.257 s^{0.195}    
\end{equation}
instead of the previous adjustment to UA5 data:
\begin{equation}
  ~~<dN_{ch}/d\eta> = 0.74s^{0.105}
  <N_{ch}>       = 7.2 s^{0.127} - 7
\end{equation}
Those asympotic tendencies turn to be an intermediate between the behaviour predicted by the Landau hydrodynamical model (having overestimated the conversion in an $s^{0.25}$ multiplicity of the major part of the CM'S available energy) and  the prediction of Feynman's scaling in favour of $s^{0.13}$.

Our collision Monte Carlo generator has provided in those conditions the pseudo rapidity distributions $dN/d\eta$ for charged and neutral particles, as shown for charged hadrons on fig.5.
 \begin{figure}                                                                 
 \includegraphics[width=0.5\textwidth,clip]{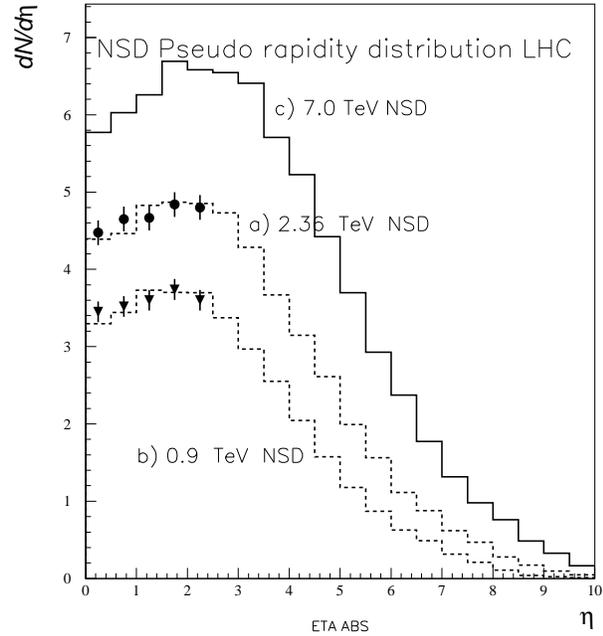} \hfill                   
 \caption{Charged pseudo rapidity distributions for the inclusive NSD component averaged over 3000 collisions}   
  \end{figure}    
The central pseudorapidityies are in agreement with CMS, ATLAS and ALICE measurements. Other characteristics of the p-p collisions are shown in Table~\ref{l2ea4-t1}.  
The inclusive and semi-inclusive data are compared in Fig.~2 at $\sqrt(s) = 1.37$ TeV (comparison with JF1aF1). 
The discrepancy with earliest extrapolations based on UA5 data are not visible at this energy, but according to Table~\ref{l2ea4-t1} we can expect clear asymptotic tendencies such as larger multiplicities, larger violation of KNO scaling, larger central rapidity densities and a larger energy flow transferred to secondaries (as suggested by the inelasticity $K_{tot}$ in Table~\ref{l2ea4-t1}). The microscopic models with intrinsic parameters tuned to involve those effects will produce for proton initiated showers more muons and a maximum depth of the e.m. component at higher altitude. 
Consequently, the present values proposed for the primary mass are overestimated in the knee region and beyond.
%


\begin{table}[t]
\begin{center}
\caption{Characteristics of NSD inclusive data}
\begin{tabular}{|l|c|c|c|}
\hline \textbf{$\sqrt(s)$} & \textbf{$0.9$TeV} & \textbf{$2.6$TeV} &
\textbf{$7.0$TeV}
\\
\hline$<dN_{ch}/d\eta>$  & 3.48 & 4.47 & 5.95 \\
\hline $<N_{ch}>$  & 33.6 & 49.27 & 73.86 \\
\hline $<K_{tot>}$ &0.59 & 0.64 &0.65 \\
\hline $<p_{t}>$& 0.46 & 0.49 & 0.54 \\
\hline
\end{tabular}
\label{l2ea4-t1}
\end{center}
\end{table}


\subsection{Extensive air showers at UHE}
At ultra high energy,
the maximum depth $X_{max}$ is also an indicator of the characteristics of the primary interaction in reason of the relation of the individual maximum of the main subcascade with two factors ; the vertex of the earliest collisions dependent on  the inelastic cross section on one hand, the individual energy given to the most energetic $\gamma$'s according to the profile of the fragmentation region on the other hand. 
The situation is compared on fig.5 with the different models \cite{cap3} together with the data collected by HIRES and AUGER. In contrast with HIRES, above an energy of $2-3$ EeV, AUGER suggests that the penetration power of cosmic particle in atmosphere levels off with an asymptotic tendency pointing to the alone event and most energetic event of the Fly's Eye. Is here a hint of a phase transition to QGP (an energy density larger than $30$ GeV/Fm$^{3}$ can be expected for 100 EeV collisions)? Is it the signature of a transition to a heavy primary composition?  We observe that the last bins of energy in fig.6 for AUGER contains hardly 20 primaries and the hybrid showers selected by the surface detector might not reflect the actual situation of the primary flux in those bins.
 \begin{figure}                                                                  \includegraphics[width=0.5\textwidth,clip]{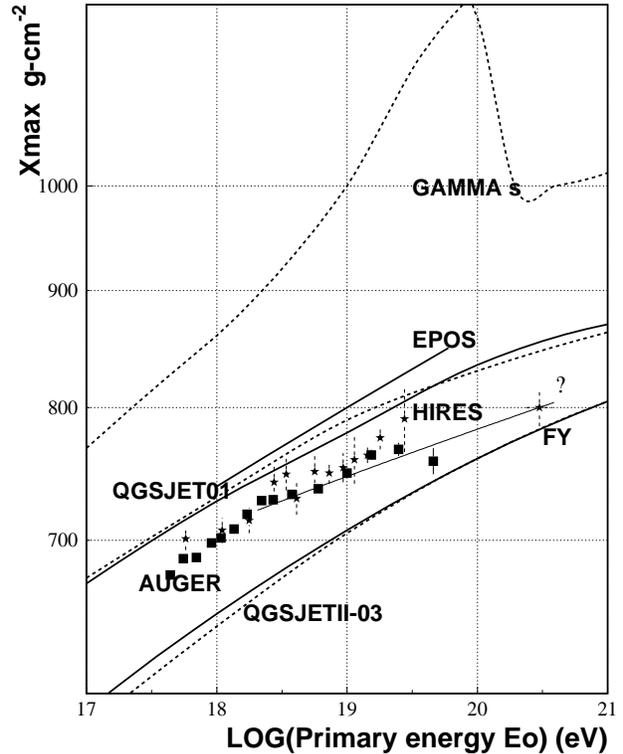} \hfill                    \caption{Penetration power of the UHE cosmic ray particles in the atmosphere.  The solid line with a question mark corresponds to the tendency suggested by AUGER data}
    \end{figure}

\section{CONCLUSION}
The preliminary data coming from the LHC implies at least an incremental revision of the models used in EAS simulation. The interpretation of the simulations will turn consequently to lower the primary mass composition at the knee level
 and above. The modest changes expected up to $10^{8}$ GeV could be more important at AUGER level and would favour the interpretation of HIRES (GZK behaviour and pure proton component).
 
 In the LHC energy range, the conjecture is similar to the situation preceding the discovery of the charm, proposed in cosmic rays, rejected in a first accelerator experiment and finally discovered in further measurements. Furthermore LHC data will help to clarify the role of complex fluctuations in XREC data: updated XREC experiments (for instance with emulsion bricks) could also be performed in parallel with the LHC in order to determine the behaviour of the most energetic secondaries in fragmentation region.   




\bigskip 


\bigskip 

\end{document}